\documentclass[letterpaper, 10 pt, conference]{ieeeconf}  

\IEEEoverridecommandlockouts                              
\usepackage{IRM}
\title{\LARGE \bf Contraction-Based Methods for Stable Identification and\\ Robust Machine Learning: a Tutorial}

\author{Ian R. Manchester, Max Revay, Ruigang Wang  
\thanks{This work was supported by the Australian Research Council.}
\thanks{The authors are with the Australian Centre for Field Robotics and Sydney Institute for Robotics and Intelligent Systems, The University of Sydney, Sydney, NSW 2006, Australia (e-mail: {\tt\small  ian.manchester@sydney.edu.au}).}%
}

\begin{document}

\maketitle
\thispagestyle{empty}
\pagestyle{empty}

\begin{abstract}
This tutorial paper provides an introduction to recently developed tools for machine learning, especially learning dynamical systems (system identification), with stability and robustness constraints. The main ideas are drawn from contraction analysis and robust control, but adapted to problems in which large-scale models can be learnt with behavioural guarantees. We illustrate the methods with applications in  robust image recognition and system identification.
\end{abstract}

\section{Introduction}
The purpose of this tutorial is to provide a gentle introduction to some recently developed tools for robust system identification and machine learning, building upon ideas from robust control \cite{zhou1996robust} and contraction analysis \cite{Lohmiller:1998}. In this tutorial we gradually build from the most basic case of linear time-invariant systems to the latest  results on complex nonlinear models incorporating equilibrium neural networks.

The main aim of these methods is to construct model parameterizations that \textit{guarantee} behavioural properties of the model such as stability and robustness. Such models can then be fit to data in various ways. One hopes that the data are representative and an accurate model is the outcome, but necessarily incomplete data mean that black-box models can often exhibit undesirable or non-physical behaviours. Introducing model behavioural constraints help to ensure that \textit{whatever the case}, the resulting model is well-behaved, and also serves as a form of model regularization.

The key ideas were developed over a series of papers including \cite{tobenkinConvexOptimizationIdentification2010, tobenkinConvexParameterizationsFidelity2017, umenbergerSpecializedInteriorPointAlgorithm2018, umenbergerMaximumLikelihoodIdentification2018, umenbergerConvexBoundsEquation2019, revay2020contracting, revay2020lipschitz, revay2020convex, revay2021recurrent}, however model structures, assumptions, and notation varied somewhat across these papers so our aim in this tutorial is to provide a simple and consistent explanation of the central ideas.

\subsection{Related Literature}

System identification, i.e. learning dynamical systems that can reproduce observed input/output data is a common problem in the sciences and engineering, and a wide range of model classes have been developed, including finite impulse response models \cite{schetzen20016volterra} and models that contain feedback, e.g. nonlinear state-space models \cite{schon2011system}, autoregressive models \cite{billings2013nonlinear}  and recurrent neural networks \cite{mandic2001recurrent}.

When learning models with feedback it is not uncommon for the model to be unstable even if the data-generating system is stable, and this has led to a large volume of research on guaranteeing model stability. Even in the case of linear models the problem is complicated by the fact that the set of stable matrices is non-convex, and various methods have been proposed to guarantee stability via regularization and constrained optimization \cite{maciejowskiGuaranteedStabilitySubspace1995, van2001identification, lacy2003subspace, Nallasivam11, miller2013subspace, umenbergerConvexBoundsEquation2019, mamakoukasMemoryEfficientLearningStable2020}. 

For nonlinear models, there has also been a substantial volume of research on stability guarantees, e.g. for polynomial models \cite{tobenkinConvexOptimizationIdentification2010, tobenkinConvexParameterizationsFidelity2017, umenbergerSpecializedInteriorPointAlgorithm2018}, Gaussian mixture models \cite{khansari-zadehLearningStableNonlinear2011a}, and recurrent neural networks \cite{miller2018stable, revay2020contracting, revay2020convex}, however the problem is substantially more complex than the linear case as there are many different definitions of nonlinear stability and even verification of stability of a given model is challenging. Contraction is a strong form of nonlinear stability \cite{Lohmiller:1998}, which is particularly well-suited to problems in learning and system identification since it guarantees stability of \textit{all} solutions of the model, irrespective of inputs or initial conditions. This is important in learning since the purpose of a model is to simulate responses to previously unseen inputs. The first method for learning contracting nonlinear models was given in \cite{tobenkinConvexOptimizationIdentification2010}, and contraction constraints were also used in \cite{tobenkinConvexParameterizationsFidelity2017, umenbergerMaximumLikelihoodIdentification2018, umenbergerSpecializedInteriorPointAlgorithm2018, sindhwani2018learning, miller2018stable, revay2020contracting, revay2020convex}.

More generally, a variety of behavioural constraints have  been investigated in the literature: in \cite{revay2021Distributed} network models with contraction and monotonicity properties are identified,  while \cite{manchester2011identification} finds \textit{transverse contracting} \cite{manchester2014transverse} models that can exhibit limit cycles, and \cite{singh2020learning} finds stabilizable models via control contraction metrics \cite{Manchester:2017}. 

An important behavioural constraint is model \textit{robustness}, characterised in terms of sensitivity to small perturbations in the input. It has recently been shown that recurrent neural network models can be extremely fragile \cite{cheng2020seq2sick}, i.e. small changes to the input produce dramatic changes in the output.

Formally, sensitivity and robustness can be quantified via \textit{Lipschitz bounds} on the input-output mapping defined by the model, e.g. incremental $\ell_2$
gain bounds. In
machine learning, Lipschitz constants are used in the proofs of
generalization bounds \cite{Bartlett:2017} and
guarantees of robustness to adversarial attacks \cite{Huster:2018,Qian:2019}. 
There is also ample empirical evidence to suggest that Lipschitz regularity (and model stability) improves generalization in machine learning \cite{gouk2021regularisation}, system identification \cite{revay2020convex, revay2021recurrent} and reinforcement learning \cite{russo2019optimal}.

Unfortunately, even calculation of the Lipschitz constant of a feedforward (static) neural networks is NP-hard \cite{virmax2018lipschitz} and instead approximate bounds must be used. The tightest bound known to date is found by using quadratic constraints to construct a behavioural description of the neural network activation functions \cite{Fazlyab:2019}. 
Extending this approach to training new neural networks with a prescribed Lipschitz bound is complicated by the fact that the model parameters and IQC multipliers are not jointly convex. In \cite{pauli2021training}, Lipschitz bounded feedforward models were trained using the Alternating Direction Method of Multipliers, although this added significant computational cost compared to unconstrained training. In \cite{revay2020lipschitz} a parameterisation was given allowing equivalent constraints to be satisfied via unconstrained optimization.

In this paper, we give an introduction to a family of tricks and techniques that has proven useful generating convex sets of linear and nonlinear dynamical models that incorporate various stability and robustness constraints.

\section{Stability and Contraction}

Given a dataset $\tilde{z}$, we consider the problem of learning a nonlinear state-space dynamical model of the form
\begin{equation}\label{eq:rnn}
	x_{t+1}=f(x_t,u_t,\theta), \quad y_t=g(x_t,u_t,\theta) 
\end{equation}
that minimizes some loss or cost function depending (in part) on the data, i.e. to solve a problem of the form
\begin{equation}\label{eq:learning}
	\min_{\theta\in \Theta} \; \mathcal{L}(\tilde{z},\theta).
\end{equation}
In the above, $x_t\in\R^n,u_t\in\R^m,y_t\in\R^p,\theta\in\Theta\subseteq\R^N$ are the model state, input, output and parameters, respectively. Here $ f:\R^n\times\R^m\times\Theta\rightarrow\R^n$ and $g:\R^n\times\R^m\times\Theta\rightarrow\R^p$ are piecewise continuously differentiable functions.

In the context of system identification we may have $\tilde z = (\tilde y, \tilde u)$ consisting of finite sequences of input-output measurements, and aim to minimize {\em simulation error}: 
\begin{equation}\label{eq:sim_error}
 \mathcal{L}(\tilde{z},\theta)=\|y-\tilde{y}\|_T^2 
\end{equation}
where $ y=\mathfrak{R}_a(\tilde{u}) $ is the output sequence generated by the nonlinear dynamical model \eqref{eq:rnn} with initial condition $x_0=a$ and inputs $u_t=\tilde{u}_t$. Here the initial condition $a$ may be part of the data $\tilde z$, or considered a learnable parameter in $\theta$.

In this paper, we are primarily concerned with constructing model parameterizations that have favourable stability and robustness properties. 
The particular form of nonlinear stability we use is the following:
\begin{defn}\label{def:contracting}
	A model \eqref{eq:rnn} is said to be \textit{contracting} if for any two initial conditions $ a, b\in\R^n $, given the same input sequence $u$, the state sequences $ x^a $ and $ x^b $ satisfy $ |x_t^a-x_t^b|\leq K\alpha^t|a-b| $ for some $K>0$ and $\alpha\in [0,1)$. 
\end{defn}

\begin{figure}
  \includegraphics[width=0.85\columnwidth]{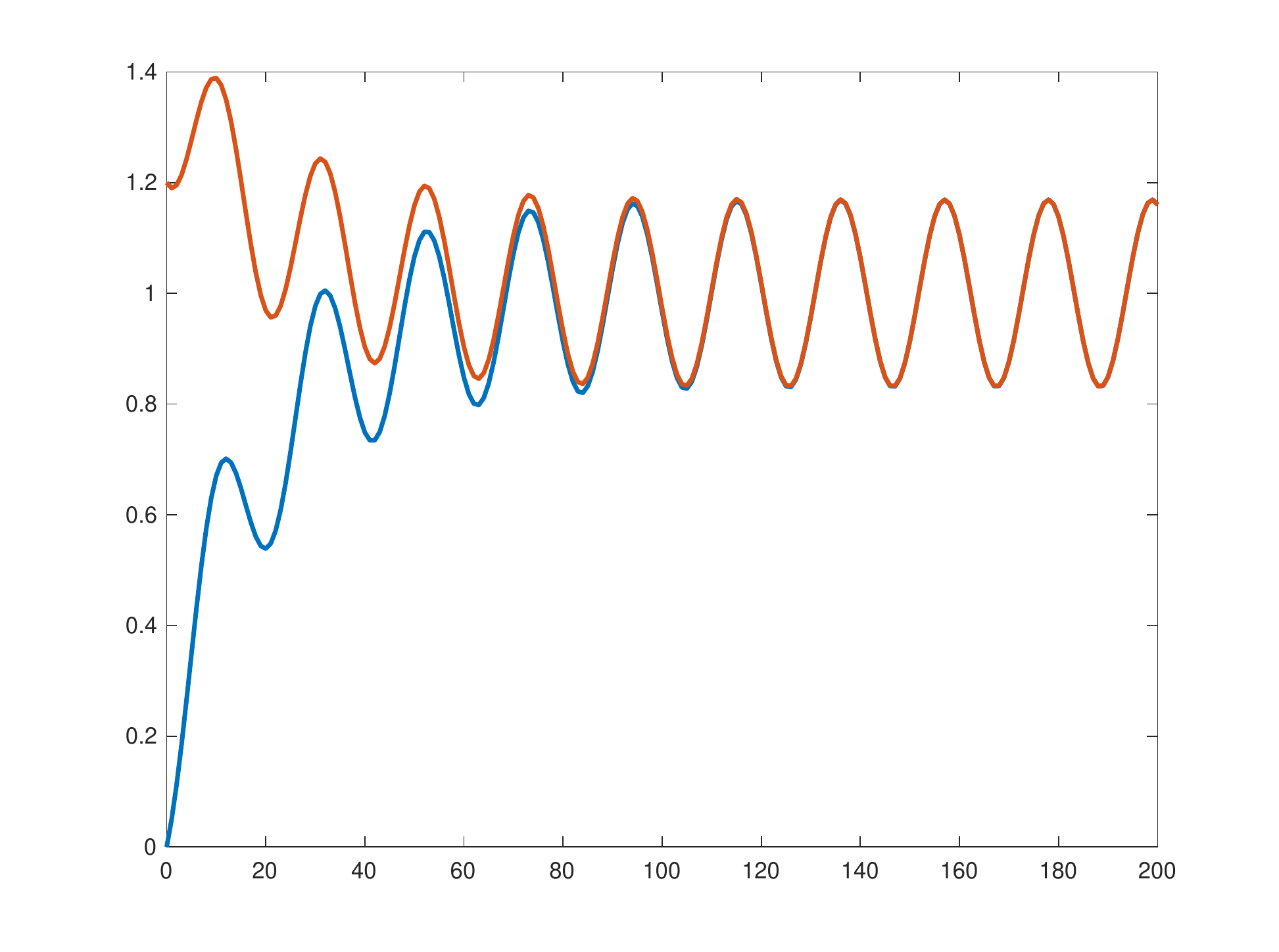}
  \caption{Contracting systems forget initial conditions and converge to the same response to arbitrary inputs.}
  \label{fig:increment}
\end{figure}

The main idea is that contracting models forget their initial conditions exponentially, as illustrated in Fig. \ref{fig:increment}. This is useful for system identification and learning since typically initial conditions are not known, and the purpose of the model is to compute responses to previously-unseen inputs, so  stability must be guaranteed for arbitrary inputs.

In general, contraction can be established by finding a \textit{contraction metric} \cite{Lohmiller:1998}. Formally, this is a Riemannian metric imbuing tangent vectors  to the state-space (roughly speaking, infinitesimal displacements) with a smoothly-varying notion of length. However, for the purposes of this tutorial we will use constant (state independent) metrics and hence can consider equivalently \textit{finite displacements}, between. We note that \cite{tobenkinConvexParameterizationsFidelity2017, umenbergerSpecializedInteriorPointAlgorithm2018, umenbergerConvexBoundsEquation2019} extend the methods in this tutorial to state-dependent Riemannian metrics.

\section{The Basic Idea}
To introduce some key concepts, we set aside contraction for a moment and consider ensuring asymptotic stability of a model of the form:
\begin{equation}\label{eq:xp_fx}
  \xp = f(x)
\end{equation}
where for notational simplicity $\xp$ is shorthand for $x({t+1})$ and $x$ denotes $x(t)$.

The aim is to learn a model and prove that the origin is the unique globally-stable equilibrium. I.e. $f(0)=0$ and $x(t)\rightarrow 0$ as $t\rightarrow \infty$ from all initial conditions $x(0)\in\R^n$.

There are several ways to define stability of such a system, but one useful definition is $\ell^2$ stability, i.e.
\begin{equation}\label{eq:x_l2_stab}
\sum_{t=0}^\infty |x(t)|^2  <\infty
\end{equation}
for solutions of \eqref{eq:xp_fx} from any initial condtion $x(0)$. Note for this sum to be finite, it must be the case that $|x(t)|\rightarrow 0$ quite fast, and in fact this type of stability can be related other strong forms such as exponential stability \cite{Megretski:1997}.

When the function $f$ defining the model is fixed and known, the standard way to verify stability is to find a \textit{Lyapunov function}:

\begin{cond}
Suppose there exists a function $V(x)$ which is non-negative: $V(x)\ge 0$ for all $x$, for which
\begin{align}\label{eq:V_f}
 V(x)-V(f(x))\ge \epsilon |x|^2 \quad \forall x.
\end{align}
Then the system \eqref{eq:xp_fx} is $l^2$ stable.
\end{cond}

This follows by simply summing the inequality over $t \in [0, T]$ for arbitrary $T\ge 0$, we have
\begin{align}
  V(x_0)-V(x_T)\ge\sum_{t=0}^T \epsilon |x|^2
\end{align}
and since $V(x_T)\ge 0$ regardless of $x_T$ and $T\ge 0$ was arbitrary, we have
$
\sum_{t=0}^\infty |x(t)|^2\le  \frac{1}{\epsilon}V(x_0)
$
so \eqref{eq:x_l2_stab} holds and the origin is globally stable.

While verifying \eqref{eq:V_f} for a known $f$ can be challenging in many cases, from the computational perspective it has the advantage that it a convex (in fact linear) condition on the function $V$. For example, if candidate functions $V$ are linearly-parameterized then this implies a convex constraint on the parameters.

On the other hand, if the objective is to \textit{search} for $f$, as it is in system identification, then we strike a problem: the term $V(f(x))$ is not jointly-convex in $V$ and $f$. A key focus of this tutorial is on how to disentangle these functions in the stability condition.

Now, an equivalent statement to \eqref{eq:V_f} is the following: 

\begin{cond}
For all pairs $x, \xp \in \R^n$ such that $\xp = f(x)$, the following inequality holds:
\begin{equation}\label{eq:V_xp}
V(x)-V(\xp)\ge \epsilon |x|^2.
\end{equation}
\end{cond}
However, such conditional statements are hard to verify due to the complex interdependence of the functions $V$ and $f$ and the pairs $x, \xp$ on which the condition holds.

A common strategy is to replace such a conditional statement with an unconditional statement via the introduction of \textit{multipliers}:

\begin{cond}
There exists a \textit{multiplier} vector $\lambda(x, \xp)$ such that for all pairs $x, \xp \in \mathbb R^n$.
\begin{equation}\label{eq:sproc1}
V(x)-V(\xp)+ \lambda(x,\xp)\trn(\xp-f(x))\ge \epsilon |x|^2.
\end{equation}	
\end{cond}
In robust control this kind of approach is usually called the \textit{S-Procedure}, and is closely related to \textit{positivstellensatz} constructions in semialgebraic geometry \cite{Parrilo:2003}. 
 
Here we have made some progress, in that the above condition is now \textit{linear} in the unknowns $V$ and $f$. However, there are still two issues:
 
   It is well known that conditions such as \eqref{eq:sproc1} are generally very conservative if the multiplier $\lambda$ is fixed, but condition \eqref{eq:sproc1} is \textit{not} jointly convex in $\lambda$ and $f$, since their product $\lambda f$ appears.
   Even if one could search for $\lambda$, this may still be conservative. It is clear that if $(\xp-f(x))=0$ then so does $(\xp-f(x))^q$ for any integer $q$, so these could be added with additional multipliers. But these conditions are nonlinear in the function $f$.

These issues lead us to the final step in our construction. Instead of an explicit model defined by a function $f$, we search for an implicit equation 

\begin{equation}
  m(x, \xp)=0
\end{equation}
 with the property that for every $x$ there is a unique $\xp$ satisfying $m(x, \xp)=0$. Roughly speaking, since this representation is non-unique it allows some of the required flexibility of the multiplier to be ``absorbed'' into the model equations. 

This representation admits standard-form difference equations by $m(x,\xp)=\xp-f(x)$, but is more flexible: e.g. for any invertible (bijective) function $h$, the model functions 
   $m(x,\xp) = \xp-f(x)$,
   $m(x,\xp) = h(\xp-f(x))$, and
   $m(x,\xp) = h(\xp)-h(f(x))$
all define the same dynamical system. Hence we call $m$ a \textit{redundant} representation.

This brings us to our final condition:
\begin{cond}
Suppose there exists a non-negative function $V$ such that
\begin{align}
  V(x)-V(\xp)+\lambda(x,\xp)\trn m(x,\xp)\ge \epsilon |x|^2.
\end{align}
Then for all trajectories satisfying $m(x,\xp)=0$ satisfy \eqref{eq:V_xp}.
\end{cond}
Note that this condition is jointly convex in $V$ and $m$. It is still not jointly convex in $m$ and $\lambda$, but as we will see below the additional flexibility in $m$ ameliorates this problem quite significantly.

\section{Learning Stable Linear Systems}
Linear dynamical systems are by far the most comprehensively studied when it comes to questions of stability, and this extends to the issue of model stability in system identification. There are several possible approaches to guaranteeing stability of identified linear models, including state-space and transfer-function representations, but the approach we describe here has the advantage of extending naturally to the nonlinear model structures in the following sections.

A linear state-space model is defined by four matrix variables $A, B, C, D$ as below:
\begin{align}
  \xp &=Ax+Bu, \label{eq:lti_x}\\
  y&=Cx+Du.
\end{align}
This system is asymptotically -- and indeed exponentially -- stable if and only if the spectral radius (magnitude of the largest eigenvalue) of $A$ is strictly less than one. 

Despite the simplicity of linear system stability, there is one complicating factor when trying to \textit{identify} the dynamics: the set of stable $A$ matrices is not convex. This fact is illustrated by the following simple example:
\begin{ex}
	Consider two matrices $A_a$ and $A_b$ below and their average $A_c$:
\begin{align}
  A_a = \bm{0.5&2\\0&0},\quad  A_b = \bm{0&0\\2&0.5}\\ A_c = \frac{1}{2}(A_a+A_b)= \bm{0.25&1\\1&0.25}.
\end{align}
Both $A_a$ and $A_b$ have spectral radius 0.5, so they are stable, but their average $A_c$ has spectral radius 1.25, so it is unstable.
\end{ex}

Stability of a linear system can also be established via a quadratic Lyapunov function $V(x) = x\trn Px$, leading to the following well-known condition:

\begin{cond}
The system \eqref{eq:lti_x} is stable if and only if there exists a $P=P\trn\succ 0$ such that
\[
A\trn P A - P\prec 0.
\]
\end{cond}

It is somewhat obvious but worth noting that \textit{differences} between of solutions of \eqref{eq:lti_x} obey the linear  dynamics
\begin{equation}
  \Dxp = A\Dx
\end{equation}
where $\Dx = x^a_t-x^b_t$ and $\Dxp = x^a_{t+1}-x^b_{t+1}$, and $x^a, x^b$ are two solutions of \eqref{eq:lti_x} with different initial conditions but the same input $u$. Hence for linear systems, stability of the origin of the unforced system and incremental stability of all solutions (with inputs) are equivalent. This is not true for nonlinear systems.

Now, we have seen that the set of stable models parameterized via $A$ is non-convex, but following from the discussion in the previous section we can guess it may be beneficial to introduce an implicit representation. We introduce the new structure:
\begin{align}
  E\xp &=Fx+Ku, \label{eq:lti_implicit}\\
  y&=Cx+Du,
\end{align}
which, so long as $E$ is invertible is an equivalent to \eqref{eq:lti_x} via $A=E^{-1}F, B= E^{-1}K$. Hence this representation does not add any fundamental expressivity to the model set, but we will see below that it does allow us to convexify the set of stable models.
In this representation, differences between solutions obey the dynamics:
\begin{equation}
  E\Dxp = F\Dx
\end{equation}
Now, following the main idea from the previous section, we consider the contraction metric $\Dx^TP\Dx$, and then combine the stability condition with the implicit model equations:
\begin{align}\label{eq:lin_stab_mult}
  \underbrace{\Dxp\trn P\Dxp-\Dx\trn P\Dx}_{\textrm{Contraction metric decrease}}-\underbrace{2\Dxp\trn (E\Dxp-F\Dx)}_{=0 \textrm{ by definition of model}}\le -\epsilon |\Dx|^2.
\end{align}
Since the term $E\Dxp-F\Dx$ vanishes along model solutions, we have
\begin{align}
\Dxp\trn P\Dxp\le \Dx'P\Dx-\epsilon|\Dx|^2
\end{align}
and summing over time, we obtain for all pairs of solutions
\begin{equation}
  \sum_{t=0}^\infty |x^a_t-x^b_t|^2 \le \frac{1}{\epsilon} (x^a_0-x^b_0)\trn P(x^a_0-x^b_0)
\end{equation}
I.e. the system is stable.

The key benefit of this formulation is that condition \eqref{eq:lin_stab_mult} is convex in the model parameters $E, F, P$, and can be rewritten as the linear matrix inequality (LMI):
\begin{equation}\label{eq:lin_lmi}
  \bm{(E+E^T-P)&-F\\-F^T &P}\succ 0.
\end{equation}
Furthermore, for \textit{every} stable linear system \eqref{eq:lti_x}, there exists a representation in the implicit form \eqref{eq:lti_implicit} such that \eqref{eq:lin_lmi} holds \cite{tobenkinConvexOptimizationIdentification2010, tobenkinConvexParameterizationsFidelity2017}. Thus we have a convex parameterization of all stable linear models.

We also note that the lower-left block of \eqref{eq:lin_lmi} ensures $P\succ 0$ and the upper-left block ensures that $E+E^T\succ 0$, which implies that all eigenvalues of $E$ have positive real parts, hence $E$ is invertible, so the model is well-posed.

\begin{remark}
	In \cite{umenbergerMaximumLikelihoodIdentification2018} this formulation was used to learn maximum likelihood models from input-output data via an expectatation maximization algorithm.
\end{remark}

\begin{remark}\label{ref:rem_direct}
	Condition \eqref{eq:lin_lmi} takes the form of an LMI, but can be used to defines a \textit{direct} (unconstrained) parameterization of stable linear models. Indeed, take $V\in\mathbb R^{2n\times 2n}$, where $n$ is the dimension of the system. Then let $H= VV^T+\epsilon I$ for small $\epsilon>0$, so $H$ is positive definite. Then partition $H$ into $n\times n$ blocks, and we can extract $P= H_{22}, F=-H_{12}, E=\frac{1}{2}(H_{11}+H_{22}+S)$ where $S$ is any skew-symmetric matrix, and the resulting model satisfies \eqref{eq:lin_lmi}. This can be useful since there are many algorithms for unconstrained optimization that are well-suited to machine learning with large data sets, e.g. stochastic gradient descent and Adam \cite{Kingma:2017}.
\end{remark}

\section{Learning Contracting Recurrent Neural Networks}\label{sec:rnn}
Next we adapt the main ideas to learning a class of nonlinear model incorporating neural networks. For nonlinear models, Lyapunov stability and contraction can be very different, e.g. a double pendulum has energy as a (non-strict) Lyapunov function, and has bounded solutions, but exhibits chaotic response to large initial conditions. Similarly, if one takes a standard recurrent neural network (RNN) \cite{mandic2001recurrent} with sigmoid activation functions, then all solutions are bounded since the activation maps to a bounded interval, but within this interval solutions can be chaotic, i.e. nearby trajectories rapidly diverge.

Contraction \cite{Lohmiller:1998} ensures that trajectories with the same input but different initial conditions always converge, and rules out chaotic behaviour, and is hence useful in learning predictive models since initial conditions are rarely known and it is the input-output response that is typically being modelled.

RNNs come in various flavours but they are generally characterized by interconnections of ``learnable'' linear mappings (weight matrices) and fixed nonlinear mappings ("activation functions"). In this tutorial, we first consider the class introduced in \cite{revay2020convex}:
\begin{align}
  \xp &= Ax+B_1w+B_2u,\notag
 \quad v = C_1x+D_{12}u\\
  	y &= C_2x +D_{21}w+D_{22}u, \notag
\quad	w =\sigma(v) 
\end{align}
where $\sigma$ is a scalar nonlinearity applied elementwise. We assume that it is slope-restricted in $[0, 1]$, i.e.
	\begin{equation}\label{eq:slope}
	0\leq \frac{\sigma(v^a)-\sigma(v^b)}{v^a-v^b}\leq 1, \quad \forall v^a,v^b\in\R,\; v^a\neq v^b.
	\end{equation}
	All widely used activation functions (ReLU, leaky ReLU, sigmoid, tanh) satisfy this restriction, possibly after rescaling.
	
This model class includes many commonly-used model structures, including classical RNNs \cite{mandic2001recurrent}, linear dynamical systems, and single-hidden-layer feedforward neural networks, see \cite{revay2020convex} for details.

To establish that a model of this form is contracting we will need to deal with the nonlinearity $\sigma$. We will do this by introducing some quadratic bounds for the signal. 
 With $w=\sigma(v)$ the slope restriction can be rewritten as 
\[(v_a-v_b)(w_a-w_b) \ge (w_a-w_b)^2\]
 Since this holds for \textit{all} activation channels (the elements of $v$), we can also take positive combinations of these:
\begin{align}
  \sum_i\lambda_i[(v_a^i-v_b^i)(w_a^i-w_b^i) - (w_a^i-w_b^i)^2]\notag\\
  =(C_1\Dx-\Dw)\trn\Lambda\Dw \ge 0
\end{align}
where $\Lambda$ is a diagonal matrix, with diagonal elements $\lambda_i>0$.

This quadratic constraint is however not jointly convex in the model parameter $C_1$ and the multiplier matrix $\Lambda$. The trick is to absorb the multiplier into model by defining $\tilde C = \Lambda C_1$, obtaining the equivalent constraint:
\[ (\tilde C\Dx-\Lambda\Dw)\trn\Dw\ge 0.\]
And we note that since $\Lambda\succ 0$, $C_1$ is recoverable from $\tilde C$.

%
%
%
%
%
%
Then, by following much the reasoning from the previous section, we construct the following constraint:
\begin{align}\label{eq:rnn_contraction}
 & \underbrace{\Dxp\trn P\Dxp-\Dx\trn P\Dx}_{\textrm{Lyapunov function decrease}}\notag\\-&\underbrace{2\Dxp\trn (E\Dxp-F\Dx-B_1\Dw)}_{=0 \textrm{ due to linear block}}\notag\\+&\underbrace{2(\tilde C\Dx-\Lambda\Dw)\trn\Dw}_{\ge 0 \textrm{ due to sector condition}}\le -\epsilon |\Dx|^2
\end{align}
As depicted, the term incorporating the linear dynamics vanishes, while the term incorporating the sector condition is always non-negative, and hence can be dropped without affecting the truth of the inequality. So for all pairs of model trajectories we have
\begin{align}
\Dxp\trn P\Dxp\le \Dx\trn P\Dx-\epsilon|\Dx|^2
\end{align}
and hence the system is contracting.

Now, as in the previous section \eqref{eq:rnn_contraction} is convex in the model parameters and can be written as an LMI:	
\begin{equation}\label{eq:rnn_contraction}
  \bm{(E+E\trn -P)&-F&-B_1\\
   -F\trn & P &  -\tilde C\trn\\
  -B_1\trn & -\tilde C &2\Lambda }\succ 0.
\end{equation}
Hence we have constructed a convex parameterization of contracting recurrent neural networks. Furthermore, by examining the upper-left two-thirds and comparing to \eqref{eq:lin_lmi}, it is clear that if $B=\tilde C=0$ these are equivalent, so the set of contracting RNNs contains all stable linear models as a subset. This is a natural requirement, but it is \textit{not} satisfied by previously-proposed sets of contracting RNNs, e.g. \cite{miller2018stable, revay2020contracting}.
%

\section{Robust Recurrent Neural Networks}\label{sec:robust_rnn}
 Beyond contraction, it can also be important to ensure a model is \textit{robust}, i.e. not highly sensitive to small changes in the input. One measure for this is the $\ell_2$ Lipschitz constant of the input-output map, a.k.a. the incremental $\ell^2$ gain.
 
 A model satisfies a Lipschitz bound of $\gamma$ if, for all pairs of inputs $u^a, u^b$, corresponding solutions (from identical initial conditions) satisfy
\begin{equation}\label{eq:inc_l2}
  \sum_{t=0}^T|y^a_t-y^b_t|^2\le \gamma^2 \sum_{t=0}^T|u^a_t-u^b_t|^2
\end{equation}
for all $T>0$.
To obtain such bounds, we use incremental form of the model:
\begin{align*}
  \Dxp &= A\Dx+B_1\Dw,\\
  \Delta_v &= C_1\Dx+D_{12}\Delta_u,\\
  	\Delta_y &= C_2\Dx +D_{21}\Delta_w.
\end{align*}
Now, following the same procedure as above we can verify the Lipschitz bound via an incremental dissipation inequality:
\begin{align}
&  \underbrace{\Dxp\trn P\Dxp-\Dx\trn P\Dx}_{\textrm{Storage function decrease}}\notag\\-&\underbrace{2\Dxp\trn (E\Dxp-F\Dx-B_1\Dw-B_2\Delta_u)}_{=0 \textrm{ due to linear block}}\notag\\&+\underbrace{2(\tilde C\Dx+\tilde D\Delta_w-\Lambda\Dw)\trn\Dw}_{\ge 0 \textrm{ due to sector condition}}\le  \underbrace{- |\Delta_y|^2+\gamma^2|\Delta_u|^2}_{l^2 \textrm{ gain} }\label{eq:l2gaincond}
\end{align}
with $\tilde D = \Lambda D_{21}$.
this is jointly convex in $P,E,F,B,\tilde C, \tilde D, \Lambda$ and, but summing over $t$ we have
\[
\sum_{t=0}^T[- |\Delta_y|^2+\gamma^2|\Delta_u|^2]\ge \Delta_T^TP\Delta_T-\Delta_0^TP\Delta_0.
\]
With identical intitial conditions, $\Delta_0=0$, and since $P\succ 0$, $\Delta_T^TP\Delta_T\ge 0$. So we have
\[
\sum_{t=0}^T[- |\Delta_y|^2+\gamma^2|\Delta_u|^2]\ge 0
\]
which can be rearranged to obtain \eqref{eq:inc_l2}. Once again, \eqref{eq:l2gaincond} is convex in the model parameters, and can be written as an LMI (see \cite{revay2020convex}).

\section{Equilibrium Networks}\label{sec:lben}
 The RNN structure in the previous two subsections incorporated single-hidden-layer neural networks, but much of the recent interest in machine learning has focused on \textit{deep} (multi-layer) neural networks, and variations with ``implicit depth'' such as \textit{equilibrium networks} \cite{winston2020monotone, revay2020lipschitz}, a.k.a. \textit{implicit deep models} \cite{ghaoui2019implicit} have recently attracted interest.
   
In this section, we consider nonlinear static maps defined by equilibrium networks, but we will keep the notation consistent with the previous sections. We can modify the nonlinearity by incorporating non-zero $D_{11}$ below:
\begin{equation}\label{eq:implicit}
   w=\sigma(D_{11}w+D_{12}u+b_w), \quad y=D_{21}w+b_y
\end{equation}
These have been called equilibrium networks because they can be seen as computing equilibria of difference or differential equations:
$w_{t+1}=\sigma(D_{11}w+D_{12}u+b_z)$, or $\dot w = -w+\sigma(D_{11}w+D_{12}u+b_z)$. Equilibrium networks are quite flexible, e.g. consider an $L$-layer feedforward network, described by the following recursive equation
\begin{equation}\label{eq:feedforward}
	\begin{split}
    z^0=u,\;
	z^{\ell+1}=\sigma(W^\ell z^\ell+b^\ell),\;
	y=W^Lz^L+b^L
  \end{split}
\end{equation}
with $ \ell=0,\ldots,L-1$.
 This can be rewirtten as an equilibrium network (\ref{eq:implicit}) with hidden units $ z=\mathrm{col}(z^1,\ldots,z^L)$ and weights $ D_{21}=\begin{bmatrix}
  0 & \cdots & 0 & W^L
\end{bmatrix}$, 
\begin{equation}\label{eq:forward-weight}
  D_{11}=\begin{bmatrix}
    0 & & &\\
    W^1 & \ddots & & \\
    \vdots & \ddots & 0 & \\
    0 & \cdots & W^{L-1} & 0 
  \end{bmatrix},\; D_{12}=\begin{bmatrix}
     W^0 \\ 0 \\ \vdots \\ 0
  \end{bmatrix}.
\end{equation}
An equilibrium network can be thought of as an algebraic interconnection of a linear and nonlinear components:
  \[
  v = D_{11}w+D_{12}u+b_w, \quad y = D_{21}w+b_y, \quad w=\sigma(v)
  \]
  where the linear part satisfies the incremental relations:
  \[
  \Delta_v = D_{11}\Delta_w+D_{12}\Delta_u, \quad \Delta_y=D_{21}\Delta_w
  \]
  and the nonlinear part satisfies the incremental sector condition
$
  (\Delta_v-\Delta_w)^T\Lambda\Delta_w\geq 0
$
  for any positive diagonal $\Lambda$, as per Section \ref{sec:rnn}.

Since an equilibrium network is defined by an implicit equation, it may not have a solution or it may have many. We make the following definition:
\begin{defn}
	An equilibrium network is \textit{Well-posed} if a unique solution $w$ exists for all inputs $u$ and biases.
\end{defn}

Well-posedness depends on the particular weight matrix $D_{11}$. In \cite{revay2020lipschitz} it was shown that if there exists a positive diagonal matrix $\Lambda$ such that
\begin{equation}\label{eq:eq_wp}
  2\Lambda-\Lambda D_{11}-D_{11}^T\Lambda \succ 0
\end{equation}
then the equilibrium network is well-posed. 

Indeed, supposing a solution exists its uniqueness follows from similar reasoning to the previous sections. Suppose two solutions $w_1$ and $w_2$ exist with the same $u$, and let $\Dw=w_1-w_2$. Then \eqref{eq:eq_wp} states that there exists a scalar $\epsilon>0$ such that
\begin{equation}
\underbrace{2(\Lambda D_{11}\Dw-\Lambda\Dw)\trn\Dw}_{\ge 0 \textrm{ due to sector condition}}\le -\epsilon |\Dw|^2
\end{equation}
for all $\Dw$. But then the sector condition, as indicated, implies the right-hand-side is non-negative, which is only possible of $\Dw=0$. Existence of solutions can also be shown via a contraction argument \cite{revay2020lipschitz}.

Moving beyond well-posedness, we can impose a Lipschitz bound $|\Delta_y|\le \gamma |\Delta_u|$ in much the same way, by changing the right-hand-side in the inequality above:
\begin{equation}
\underbrace{2(\Lambda \Delta_v-\Lambda\Dw)\trn\Dw}_{\ge 0 \textrm{ due to sector condition}}\le \gamma |\Delta_u|_2^2-\frac{1}{\gamma}|\Delta_y|_2^2
\end{equation}
where
$ \Delta_v = D_{11}\Delta_w+D_{12}\Delta_u$ and $\Delta_y=D_{21}\Delta_w$. This can be rewritten as the constraint:
   \begin{equation}\label{eq:lipschitz} 
    2\Lambda-\Lambda D_{11}-D_{11}^T\Lambda-\frac{1}{\gamma}D_{21}^TD_{21}-\frac{1}{\gamma}\Lambda D_{12} D_{12}^T\Lambda\succ 0.
  \end{equation}
which is convex in the parameters $\tilde D_{11} = \Lambda D_{11}$ and $\tilde D_{12} = \Lambda D_{12}$, from $D_{11}$ and $D_{12}$ can easily be recovered since $\Lambda$ is invertible. Although we don't go into detail here, one can also construct direct parameterisations enabling unconstrained optimization, see \cite{revay2020lipschitz}. 
%
%
%
%
\subsection{Robust Image Recognition}

In this subsection we present some results from \cite{revay2020lipschitz} applying such Lipschitz bounded equilibrium networks (LBEN) to the task of image recognition on the classic MNIST character recognition data set. We compare to the MON network of \cite{winston2020monotone} and the LMT network of \cite{tsuzuku2018lipschitz}.

\begin{figure}
  \includegraphics[width=1\columnwidth]{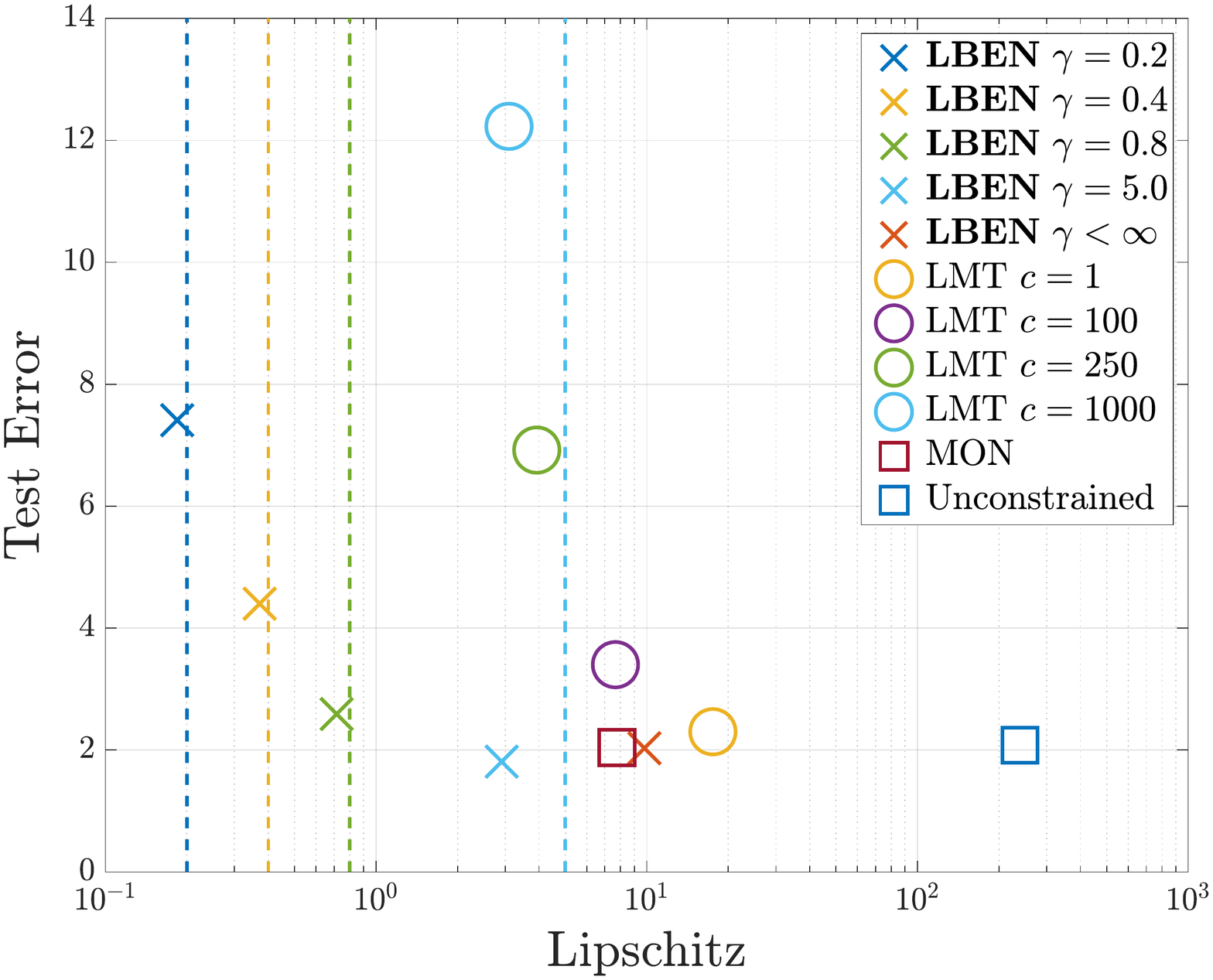}
  \caption{MNIST image recognition comparison of Lipschitz bounded equilibrium network (LBEN) to other networks: test error vs sensitivity, from \cite{revay2020lipschitz}. Vertical lines are theoretical upper bounds for LBEN models.}
  \label{fig:mnist1}
\end{figure}

Figure \ref{fig:mnist1} compares different networks in terms of test error without any perturbation (vertical axis) and a lower bound on the Lipschitz constant (horizontal axis) determined by local search for a worst-case input perturbation. The vertical lines indicate the theoretical upper bounds for Lipschitz constant for different LBENs. It can be seen that the LBEN structures allow trading off between sensitivity to perturbation and nominal test error, and that the results are better (further to the lower left) than competing methods. Furthermore, the theoretical upper bounds on the Lipschitz constant are quite close to the observed lower bounds. Note also that introducing a Lipschitz bounds actually decreases nominal test error compared to LBEN $\gamma<\infty$, which is only well-posed. I.e. we see a regularising effect.

\begin{figure}
  \includegraphics[width=1\columnwidth]{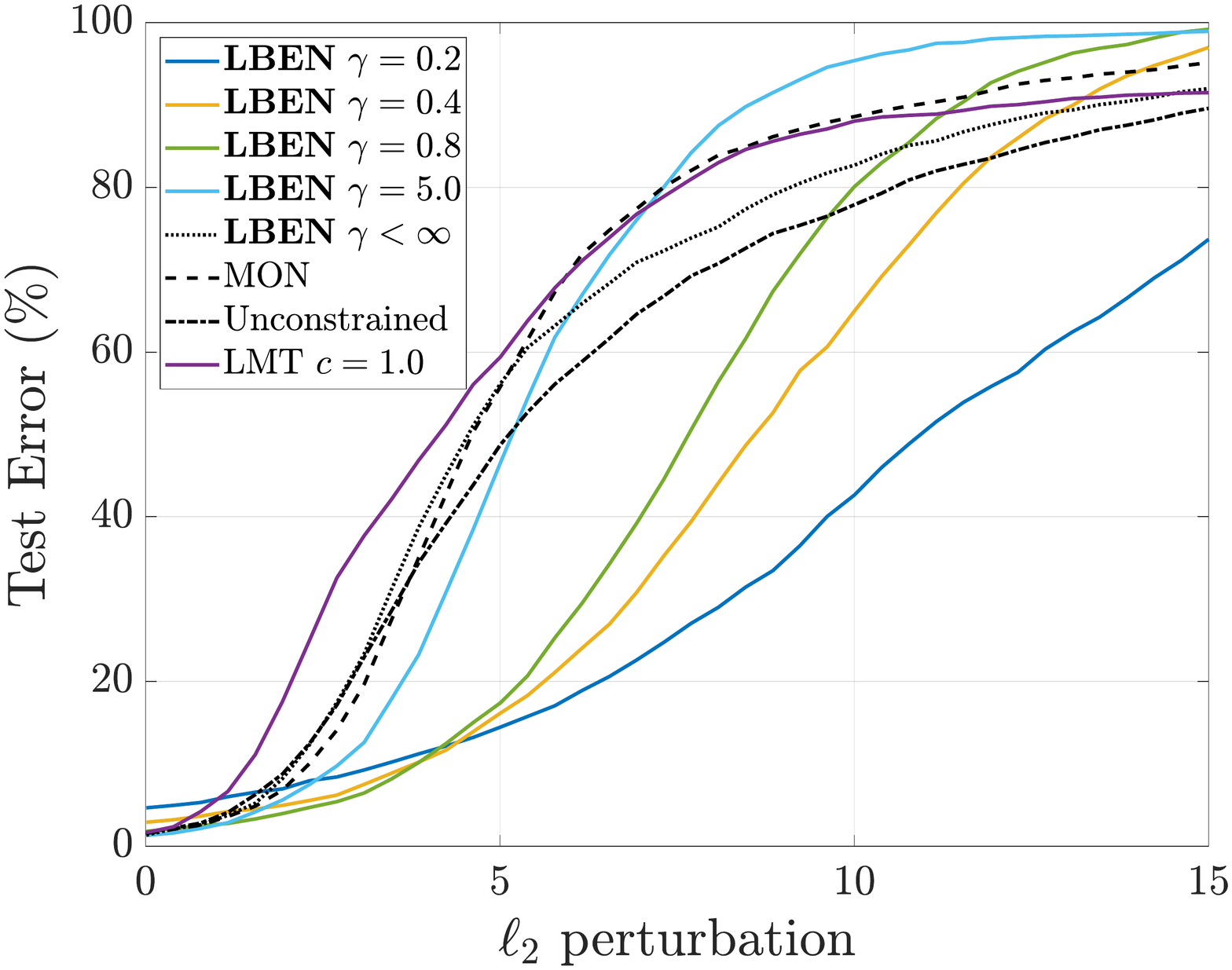}
    \caption{MNIST image recognition comparison of Lipschitz bounded equilibrium network (LBEN) to other networks: robustness to adversarial perturbations, from \cite{revay2020lipschitz}.}
    \label{fig:mnist2}
\end{figure}

In Figure \ref{fig:mnist2} we compare classification error of different networks as the size of the adversarial input increases. It can be seen that LBEN structures allow for a much slower degradation in test error with only minor affect on nominal (perturbation=0) performance. E.g. with a perturbation of 5, LBEN models can achieve an error of around 15\% while all other structures are at around 50-60\% error.

\section{Recurrent Equilibrium Networks}\label{sec:ren}
In this section we discuss the \textit{recurrent equilibrium network} (REN), recently introduced in \cite{revay2021recurrent}. Roughly speaking, it is a combination of the recurrent networks of Sections \ref{sec:rnn} and \ref{sec:robust_rnn} with the equilibrium networks of Section \ref{sec:lben}.

The REN model structure is
\begin{align}
  \xp &= Ax+B_1w+B_2u,\quad 
  v = C_1x+D_{11}w+D_{12}u,\notag\\
  	y &= C_2x +D_{21}w+D_{22}u,\quad
	w =\sigma(v). \notag
\end{align}
Compared to Sections \ref{sec:rnn} and \ref{sec:robust_rnn}, the  difference is the presence of $D_{11}$, which means that the  model incorporates an equilibrium network for $w$.

We omit details here, but following the same procedure as previous sections, we obtain the following LMI for contracting RENs:
\begin{equation}\notag
  \bm{(E+E\trn -P)&-F&-B_1\\
   -F\trn & P &  -\tilde C\trn\\
  -B_1\trn & -\tilde C & 2\Lambda- \tilde D_{11}-\tilde D_{11}^T}\succ 0, 
\end{equation}
where $\tilde D_{11} = \Lambda D_{11}$.
We briefly make some remarks about this model structure:
\begin{itemize}
  \item The lower right block is the same as \eqref{eq:eq_wp}, and hence well-posedness of the equilibrium network is \textit{automatically} satisfied via the contraction  constraint.
  \item Compared to the structure in Section \ref{sec:rnn}, the lower-right block is no longer constrained to be diagonal. Hence it is now straightforward to construct a \textit{direct} parameterization of contracting RENs, enabling learning via unconstrained optimization methods such as stochastic gradient descent. We refer the reader to \cite{revay2021recurrent} for details.
  \item We have also extended the approach of Section \ref{sec:robust_rnn} to RENs and, in fact, achieve a direct parameterization of robust (Lipschitz bounded) RENs. See \cite{revay2021recurrent} for details.
\end{itemize}

\subsection{Application to Robust System Identification}

We have applied the robust REN structure to the F16 ground vibration data set from \cite{noel2017f16}. This benchmark problem exhibits a challenging combinations of high dimension, highly resonant responses, and significant nonlinearity.

\begin{figure}
  \includegraphics[width=1\columnwidth]{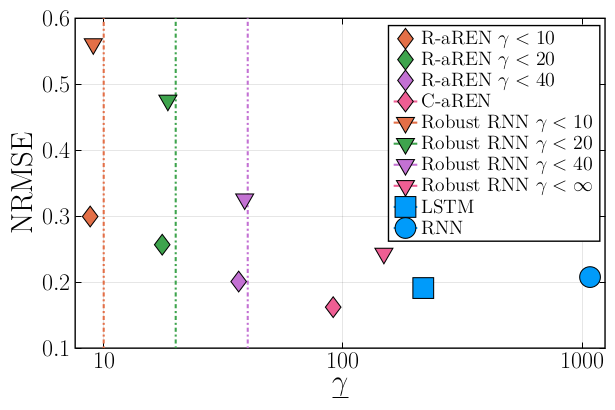}
  \caption{Comparison of different model structure on the F16 nonlinear system identification benchmark: test error vs model sensitivity. Vertical lines are theoretical upper bounds on sensitivity for REN and Robust RNN models.}
  \label{fig:F16_1}
\end{figure}

\begin{figure}
  \includegraphics[width=1\columnwidth]{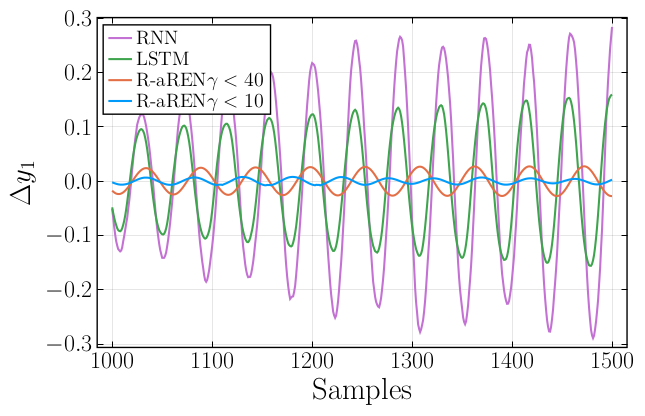}
    \caption{Zoomed in view of perturbations to trajectories on the F16 nonlinear system identification benchmark.}
    \label{fig:F16_2}
\end{figure}

In Figure \ref{fig:F16_1}, we see a plot of normalised root-mean-square test error without perturbations (vertical axis) vs sensitivity of models to perturbations. It can be seen that the REN structure provides significantly improved combination of nominal fit and robustness to perturbations compared to previous structures such as the Robust RNN from Section \ref{sec:robust_rnn} and the widely-used long short-term memory (LSTM) \cite{Hochreiter:1997} and the standard RNN structure \cite{mandic2001recurrent}. In Figure \ref{fig:F16_2} we see a zoomed-in view of the relative size of output perturbations of different models to a fixed input perturbation. it can be seen that the REN structures are much more robust to perturbation than LSTM and standard RNN.

\section{Conclusions}
This tutorial paper has given an introduction to recently developed tools for machine learning, especially learning dynamical systems (system identification), with stability and robustness constraints. 

The  main procedure is the following: (i)
   Formulate stability and robustness via contraction and incremental $\ell^2$ gain conditions.\,
  (ii) add linear model constraints to the stability/robustness conditions via Lagrange multipliers,
  (iii) add nonlinear components via incremental quadratic constraints and associated multipliers,
  (iv) make the resulting inequality convex via implicit model that structures \textit{merge} the model parameters with the multipliers.

The methods have been illustrated the methods with applications in  robust image recognition and system identification, where it was shown that significantly improved (and more robust) models can be obtained compared to existing methods. 
Our current work is in applying this approach a broader set of problems, including observer design \cite{manchester2018contracting, yi2021reduced} and reinforcement learning \cite{russo2019optimal, robertsFeedbackControllerParameterizations2011}, and investigating additional behavioural constraints, e.g. incremental passivity.

\bibliographystyle{IEEEtran}
\bibliography{REN,ref}

\end{document}